\documentstyle[aps,prd, preprint]{revtex}
\begin{document}
\draft

\title{Landau Levels in the Presence of Topological Defects}

\author{Geusa de A. Marques$^{1}$, Cl\'{a}udio Furtado$^{1,2}$, 
V. B. Bezerra$^{1}$ and Fernando Moraes$^{2}$}

\address{{$^{1}$Departamento de F\'{\i}sica - CCEN\\ Universidade Federal 
da Para\'{\i}ba,
Caixa Postal 5008, 58051-970 Jo\~{a}o Pessoa,PB,Brazil}\\
$^{2}$Laborat\'{o}rio de F\'{\i}sica Te\'{o}rica e Computacional,
Departamento de F\'{\i}sica\\ Universidade Federal de Pernambuco
50670-901 Recife, PE, Brazil\\
\parbox{14 cm}{\medskip\rm\indent
In this work we study the Landau levels in the presence of topological
defects. We analyze the behavior of electrons moving in a magnetic 
field in the presence of a continuous distribution of  disclinations,
a magnetic  screw dislocation and a dispiration. We focus on the influence 
of these topological defects on the spectrum of the electron (or hole) 
in the magnetic field in the framework of the geometric theory of defects 
in solids of Katanaev-Volovich. The presence of the defect breaks the 
degeneracy of the Landau levels in different ways depending on the defect. 
Exact expressions for energies and eigenfunctions are found for all cases. 
Using Kaluza-Klein theory we solve the Landau level problem for a dispiration
and compare the results with the ones obtained in the previous cases.}}
\maketitle

\renewcommand{\thesection}      {\Roman{section}}

\section{Introduction}

\renewcommand{\thesection}      {\arabic{section}} 

\indent

The role played by topology in the physical properties of a variety of 
systems has been a very important issue in different areas of physics as,
for example, gravitation and condensed matter physics. Topological defects 
appear in gravitation as monopoles, strings and walls\cite{VIL}. In condensed 
matter physics they are vortices in superconductors or superfluids\cite{KLi}, 
domain walls in magnetic materials\cite{KLi}, solitons in 
quasi-one-dimensional polymers\cite{hee} and dislocations or disclinations in 
disordered solids or liquid crystals\cite{Kle}. The change in the topology 
of a medium introduced by a linear defect such as a disclination, 
dislocation or dispiration in an elastic media or a cosmic defect in 
space-time produces some effects on the physical properties of the 
medium\cite{O,Val}.

In classical theory of elasticity, defects are described by the strain and 
stress tensors that contain all information about the deformation caused by 
then 
in the continuous media. Distortions were described by Volterra
\cite{Volt} in the context of elasticity theory and have later been 
subject of investigations in the context of both solid continua and crystals,
\cite{Kle,Nab}. We consider as a prototype of such an object, a hollow 
cylinder made of elastic material and cut it at a half two-plane, e.g.\ 
at $\phi = 0$ (using cylindrical coordinates $\{r, \phi, z\}$, 
the cylinder is taken to be oriented along the $z$-axis), thereby destroying
its multiple connectivity. Then take the two lips that have been
separated by the cut and translate and rotate them against each other.
Finally, after eventually removing superfluous or adding missing material, 
weld the two planes together again. This cutting and welding process is 
called Volterra process.

The influence of line defects on the electronic properties of a crystal is
an old issue in condensed matter physics\cite{BS}. Recently, alternative
approaches to study this problem have been proposed which use either a gauge 
field\cite{O,TE} or a gravity-like approach\cite{FM,BF,AA}. The modeling of
the influence of the defects on the quantum motion of electrons becomes 
reasonably easy in the latter approach, where the boundary conditions 
imposed by the defect are incorporated into
a metric. This geometric approach is based on the isomorphism that exists 
between the theory of defects in solids and  three-dimensional 
gravity\cite{KV}. From the mathematical point of view, geometric theory of 
defects in solids and the theory of gravity with torsion in the Euclidean 
formulation are the same models.  This approach is the Katanaev-Volovich 
theory of defects in solids. The advantage of this geometric description of 
defects in solids is twofold. Firstly, in contrast to the ordinary elasticity 
theory this approach provide an adequate language for continuous distribution 
of defects. Secondly, a mighty mathematical machinery of differential geometry 
clarifies and simplifies calculations. In recent years alternatives approach 
for geometric descriptions of defects have been presented by several 
authors\cite{Hol,kol,put}. 

\par

In this work we investigate the influence of topology in the study of
Landau levels in the presence of topological defects in condensed matter 
physics. In recent years, the influence of geometry in the Landau levels
has been an object of intensive research. Dunne\cite{DU} and Comtet\cite{CO} 
analyze the influence of the hyperbolic and spherical geometries in the Landau 
levels. The influence of topology in Landau levels was also investigated  
recently\cite{Fur1,F}.

In this paper we 
study the effects of different topological defects(density of disclinations, 
dislocation and dispiration) on the energy spectrum of a point charge 
(electrons or holes) in the presence of a magnetic field  parallel to  
the defect. The only approximation we make is to work in the 
continuum elastic medium, where the geometric approach makes 
sense\cite{KF,KT}. In general, defects correspond to singular curvature 
or torsion (or both) along the defect line \cite{KV}. The treatment that 
we adopt to determine Landau levels in a medium with defects is the 
following: defects are described by 
a metric $g_{ij}$  that contains all information about the deformation caused 
in the medium by them; in each case the metric is a solution of Einstein 
equation for three-dimensional media with curvature and/or torsion . 

The Hamiltonian corresponding to a charged particle in the presence of a 
vector potential in the background given by metric $g_{ij}$ is

\begin{equation}
H=\frac{1}{2m \sqrt g}(p_i-\frac{q}{c} A_i)[\sqrt g g^{ij}(p_j-\frac{q}{c} 
A_j)]\ , \label{f16}
\end{equation}
where the minimal coupling $\vec{p} \rightarrow \vec{p} - 
q/c \vec{A}$ was used and $q$ is the electrical charge of the particle.
In the literature the study of Landau levels is performed by use of 
minimal coupling of the linear momentum and potential vector,
$\vec{p} \to \vec{p}-\frac{e}{c}\vec{A}$. In this work we introduce a 
new approach that consists in the use of Kaluza-Klein~\cite{Kalu,Klei} theory 
to obtain from the geometrical point of view the Landau levels, introducing 
the magnetic field in the metric that describes the defect.

This paper is organized as follows: In sections $2$,$3$ and $4$ we
investigate the Landau levels in the presence of a density of 
disclinations, a magnetic  screw dislocation and a dispiration, respectively. 
In section $5$ we study the Landau levels in the context of Kaluza-Klein 
theory and consider the solution that corresponds to a dispiration in this
framework, and finally, in section $6$, we present the concluding remarks.

\renewcommand{\thesection}      {\Roman{section}} 

\section{Landau Levels in the Presence of a Continuous Distribution of  
Disclinations}

In a recent paper\cite{F} we have  discussed quantum motion of electrons
(or holes) under a uniform magnetic field, in the presence of a disclination.
In the geometric approach,  the medium with a disclination has the line
element given by

\begin{equation}
ds^{2} =g_{ij}dx^{i}dy^{j}= dz^{2} + d \rho^{2} + \alpha^{2}
\rho^{2} d \phi^{2}, \end{equation}

\noindent
in cylindrical coordinates. This metric is equivalent to the boundary
condition with periodicity of $2\pi\alpha$ instead of $2\pi$ around the 
z-axis. In the Volterra process\cite{Kle} of disclination creation, this 
corresponds to remove $(\alpha < 1)$ or insert $(\alpha >1)$
a wedge of material of dihedral angle $ \lambda =2 \pi( \alpha - 1)$. This 
metric corresponds to a locally flat medium with a conical singularity at the 
origin. The only nonzero components of the Riemann curvature tensor and the 
Ricci tensor are given by

\begin{eqnarray}
\label{eq:riemann}
R_{12}^{12}=R_{1}^{2}=R_{2}^{2}= 2\pi\frac{1-\alpha}{\alpha}\delta_{2}(\rho),
\end{eqnarray}
where $\delta_{2}(\rho)$ is the two-dimensional delta function in flat 
space. From expression above, it follows that if 
$0<\alpha<1 \quad  (-2\pi<\lambda<0)$ the defect carries positive curvature 
and if $1<\alpha<\infty \quad (0<\lambda<\infty)$ the defect carries 
negative curvature. This fact is very important in the curved space theory 
of amorphous solids\cite{sad} where geometrical frustration, the 
incompatibility between a given local order and the geometry of Euclidean 
space, is relieved by propagation of the local order in a space of constant 
curvature. Disclinations carrying curvature of sign opposite that of the 
curvature of the background space must be introduced\cite{KLE} in order to 
reduce the mean curvature of the model to zero, yielding a 
distorted(locally curved) structure with the desired local order, 
perforated by disclination lines: a sensible structural model for amorphous 
solids.

The study of the Landau levels in the presence of a single disclination was 
already performed\cite{F}. In this section we 
do the calculation for a continuous distribution of disclinations. The 
importance of the study of this problem in a media with a distribution of 
defects is that in a real material, generally, one has a bigger number of 
defects instead of a single one.

In a recent paper, Katanaev and Volovich\cite{kat2} considered a distribution 
of disclinations and obtained the metric which describes the continuous 
distribution of disclinations in an elastic media. We consider a circularly 
symmetric distribution of disclinations and  assume that they are uniformly 
distributed on a disk of radius R with the following density of deficit 
angles

\begin{eqnarray}
\label{deficit}
\zeta=\left\{ \begin{array}{ll} q &, \rho< R\\ 0 &, \rho>R\end{array}\right.
\end{eqnarray}

The normalized total deficit angle for this distribution is given by

\begin{eqnarray}
\label{ang}
\Theta =\frac{1}{2}qR^{2}
\end{eqnarray}

Solving three-dimensional Einstein equation, the metric that describes 
the space outside the distribution of disclinations has the form

\begin{eqnarray}
\label{metric}
ds^{2}= dz^{2} + \rho^{qR^{2}}(d\rho^{2} + \rho^{2}d\phi^{2}).
\end{eqnarray}

This means that the distribution of disclinations as seen from outside is 
the same as for one disclination with the deficit $\Theta$. Performing an
identification between the new parameter of density of disclinations 
and the old one used to describes a single disclination, we obtain that the 
line element for the density of disclinations has the same 
form of that corresponding to a single disclination, provided we do the
identification 

\begin{eqnarray}
\label{alpha}
\alpha= 1+ \frac{qR^{2}}{2}.
\end{eqnarray}

Now, doing the following change of coordinates

\begin{eqnarray}
\label{ro}
\bar{\rho}=\frac{\rho^{\alpha}}{\alpha},
\end{eqnarray}
we obtain the metric for the density of disclinations, which is given by

\begin{eqnarray}
\label{metric1}
 ds^{2}=dz^{2} + d\bar{\rho}^{2} + \alpha^{2} \bar{\rho}^{2}d\phi^{2}.
\end{eqnarray}

In order to determine the Landau levels let us consider the 
covariant Schr\"{o}dinger equation 

\begin{equation}
\label{se}
\frac{1}{2m}\nabla^{2}\Psi=i\frac{\partial\Psi}{\partial t},
\end{equation}
where the Laplacian operator is 

\begin{eqnarray}
\nabla^{2} = \frac{1}{\sqrt{g}} \partial_{i} \left( g^{ij} \sqrt{g}  
\partial_{j}\right) \hspace{0.5cm} ,
\end{eqnarray}
\noindent
and $ g = det |g_{ij}| $ stands for the determinant of the metric $g_{ij}$.

Therefore, the Schr\"odinger equation written in
the space endowed by this metric,  incorporates the boundary conditions 
dictated by the defect.

Now, let us determine the Landau levels in the presence of a distribution 
of disclinations. The configuration of the vector potential that gives a 
uniform magnetic field in conical space is 

\begin{equation}
A(\rho)=\frac{B_0\rho}{2\alpha}\hat{e}_\varphi\ . \label{f10}
\end{equation}

This expression is very similar to the Euclidean case. The unique 
difference is the presence of $\alpha$ factor in (\ref{f10}). By using of 
minimal coupling $p_i\longrightarrow p_i-\frac{e}{c}\vec{A}$, we write the 
Hamiltonian for a particle in the presence of a distribution of disclinations 
submitted to a uniform magnetic field in the $z$-direction, that is given by

\begin{eqnarray}
\hat{H}&=&-\frac{\hbar^2}{2m}\left\{\frac{\partial^2}{\partial
z^2}+\frac{1}{\rho}\left(\rho\frac{\partial}{\partial\rho}\right)+
\frac{1}{\alpha^2\rho^2}\frac{\partial^2}{\partial\phi^2}\right\}+\nonumber\\
&+&\frac{i\hbar\,q\,B}{2\alpha^2m^\ast c}\frac{\partial}{\partial\phi}+
\frac{q^2B^2_0}{8m\,c^2\alpha^2}\,  \label{f11}
\end{eqnarray}
where we have dropped the bar in variable $\rho$.

\noindent
In the limit $\alpha \longrightarrow 1$, expression (\ref{f11}) is the 
Euclidean Landau Hamiltonian. The solution of Schr\"odinger equation 
corresponding to Hamiltonian (\ref{f11}) can be written in the form

\begin{equation}
\psi(z,\rho,\phi)=R(\rho)Z(z)\Phi(\phi)\ , \label{f12}
\end{equation}
\noindent
where  $Z(z)=e^{ikj}$ and $\Phi(\phi)=e^{il\phi}$. Then, the  solution of
the radial equation is 

\begin{eqnarray}
R_{nl}=C_{nl}\exp\left(-\frac{|q|B\rho^2}{4c\hbar\alpha}\right)
\rho^{\frac{|l|}{\alpha}}\times \nonumber \\
F\left(-n,\frac{|l|}{\alpha}+1,
\frac{|q|B\rho^2}{2c\hbar\alpha^2}\right)\ , \label{f13}
\end{eqnarray}
where  $F\left(-n,\frac{|l|}{\alpha}+1,
\frac{|q|B\rho^2}{2c\hbar\alpha}\right)$ is the hypergeometric function, 
and  $C_{nl}$ is the normalization constant. 

The energy levels for this case are given by

\begin{equation}
E=\frac{\hbar \omega_{H}}{2\alpha}
\left(2n+\frac{|\ell|}{\alpha}\pm\frac{\ell}{\alpha}+1\right)+
\frac{\hbar^2k^2}{2m^\ast} \label{f14}
\end{equation}
\noindent
with  $n=0,1,2,3$ and  $\ell=0,\pm1,\pm2$, where the plus sign stands 
for holes $(q=|q|)$, the minus sign for electrons  $(q=-|q|)$ and
$\omega_{H}=\frac{|q|B}{m^\ast c}$ is the cyclotron frequency. Notice that 
the effect  of substituting an electron by a hole, $q\to -q$, is equivalent 
to inverting the sign of the $z$-component of the angular momentum , 
$\ell\to -\ell$. Since $\ell \not\in Z$, electrons and holes have the same 
energy spectrum, the same $z$ and radial wave functions and opposite 
cyclotron motions. In the limit $\alpha \to 1$, eq.(\ref{f14}) 
gives the usual Landau eigenvalues plus the kinetic energy corresponding 
to the free motion along the $z$-axis. Notice that when $\frac{1}{\alpha}=p$ 
is not integer, the conicity introduced by the defect breaks the infinite 
degeneracy of the Landau levels.

\renewcommand{\thesection}      {\Roman{section}}

\section{Landau Levels in the Presence of a Magnetic Screw Dislocation}

Dislocations are much more realistic line defects. They can modify the energy 
spectrum of electrons moving in a uniform magnetic field as reported, for 
example, by Kaner and Feldman\cite{KF} and Kosevich\cite{KT}. In these earlier 
works, Landau levels in the presence of dislocations has been 
investigated in  the context of classical theory of elasticity, via a
perturbation in the Hamiltonian. The authors  have considered  the presence 
of a screw dislocation as a delta perturbation in the Hamiltonian of 
Landau levels.

We consider an infinitely long linear  screw dislocation oriented along 
the z-axis.  The three-dimensional geometry of the medium,
in this case, is characterized by a nontrivial torsion  which is identified 
with the surface density of the Burgers vector in  the classical
theory of elasticity. In this way, the Burgers vector can be viewed as flux of 
torsion. The screw dislocation is described by the following metric, in 
cylindrical coordinates \cite{Gal},

\begin{equation}
ds^{2} =g_{ij}dx^{i}dy^{j}= \left( dz + \beta d \phi \right)^{2} + 
d \rho^{2} + \rho^{2} d \phi^{2} , \label{screw}
\end{equation}
\noindent
where $\beta$ is a parameter related to the Burgers vector $b$ 
by  $\beta = \frac{b}{2 \pi}$. This topological defect carries torsion but 
no curvature. The torsion associated to this defect corresponds to a conical 
singularity at the origin. The only nonzero component of the torsion tensor 
in this case is given by the two-form  

\begin{equation}
T^{1} = 2 \pi \beta \delta^{2} ( \rho ) d\rho \wedge d\phi   ,
\end{equation}
\noindent
where $\delta^{2}( \rho )$ is the two-dimensional delta function in flat space.

We assume that the topological defect carries in the core a magnetic 
field with magnetic flux given by $\Phi$ and outside the defect this 
magnetic field vanishes. This defect carries an internal magnetic flux and 
we call it a magnetic screw dislocation. We will consider in our calculations 
an internal magnetic field and an external uniform  magnetic field and 
analyze the consequence of these fields in the Landau levels problem, specially
in which concerns the internal magnetic field. The internal magnetic field 
is associated to an external vector potential  given by

\begin{eqnarray}
\label{int}
A_{i,\phi}=\frac{\Phi}{2\pi \rho},
\end{eqnarray}
where $\Phi$ is the magnetic flux.

Next, we turn to the calculation of the vector potential for the external 
magnetic field in the space of a  dislocation described by metric 
(\ref{screw}). We solve Maxwell equations $\vec{\nabla} \cdot \vec{B} =
0$ and $\vec{\nabla} \times \vec{B} = 4 \pi \vec{J}$ considering the 
current density  $\vec{J} = J_{0} \delta (r-R) \widehat{\phi}$, 
due to an infinite solenoid concentric with the dislocation axis. This 
ensures a uniform magnetic field $\vec{B}$ in the limit of infinite solenoid 
radius ($R \rightarrow \infty )$. As a result we find $A_{\phi} ( \rho ) = 
\frac{B \rho}{2}$ for the non-zero component of the vector potential 
$\vec{A}$. Notice that, in the non-Euclidean metric of the dislocation, the  
vector potential that produces the uniform magnetic field is identical to 
the flat space potential vector. The difference will appear in the 
differential operators, which must be defined according to the 
metric (\ref{screw}).

For a quasi-particle with effective mass $m$ in the metric of the magnetic 
 screw dislocation, the Schr\"odinger equation is 

\begin{eqnarray}
&&\left\{-\frac{\hbar^2}{2m}\left[\partial^2_z+
\frac{1}{\rho}\partial\rho(\rho\partial_\rho)+
\frac{1}{\rho^2}(\partial_\phi-\beta\partial_z +
\frac{\Phi}{2\pi})^2\right] \right. \nonumber \\
&+&\frac{iq\,B\,\hbar}{2mc}(\partial\phi-\beta\partial_z+ \frac{\Phi}{2\pi})+
\left. \frac{q^2B^2\rho^2}{8m\,c^2}\right\}\varphi=E\psi\ . \label{f17}
\end{eqnarray}
\noindent
This  equation  is solved using the ansatz

\begin{equation}
\psi(\phi,\rho,z)=Ce^{il\phi}e^{ikz}R(\rho)\ ,  \label{f18}
\end{equation}
\noindent
where $C$ is a normalization constant. Substituting this form of $\psi$ 
into Schr\"{o}dinger equation we obtain the following radial equation

\begin{eqnarray}
& & \frac{1}{\rho} \partial_{ \rho }( \rho \partial_{\rho} R(\rho)) - 
\frac{1}{\rho^{2}} ( \ell - \beta k + \frac{\Phi}{2\pi})^{2} R( \rho ) - 
\frac{q^{2}B^{2}\rho^{2}}{4 \hbar^{2}c^{2}}R(\rho) \nonumber \\
& & - \frac{qB}{2 \hbar c} ( \ell - \beta k +\frac{\Phi}{2\pi}) R( \rho ) - 
k^{2} R(\rho ) + \varepsilon R(\rho) = 0 \label{se1} ,
\end{eqnarray}

\noindent
with $\varepsilon = \frac{2mE}{\hbar^{2}}$. Now, using the change of 
variables $\sigma = \frac{\rho^{2}}{2}$, eq.(\ref{se1}) is transformed into 
\begin{eqnarray}
&&\sigma^{2} \frac{d^{2}R}{d \sigma^{2}} + \sigma \frac{dR}{d \sigma} + 
\nonumber \\
&+& \left\{ \frac{A}{2}\sigma - 
\frac{q^{2} B^{2} \sigma^{2}}{4 \hbar^{2} c^{2}} - \frac{( \ell - 
\beta +\frac{\Phi}{2\pi} )^{2}}{4} \right\} R(\sigma)=0 \  ,
\end{eqnarray}
\noindent
where $ A = \varepsilon - k^{2} - \frac{qB}{2\hbar c}(\ell - \beta k +
\frac{\Phi}{2\pi}) $. The solution of this equation is given in terms of 
the confluent hypergeometric function $F$ as

\begin{eqnarray}
R( \rho ) &=& C \exp \left\{ - \frac{m \omega_{B}}{4 \hbar} 
\rho^{2} \right\} \rho^{| \ell - \beta k + \frac{\Phi}{2\pi}|} 
\times \nonumber \\
&& F \left( - n \; , \; | \ell - \beta k + \frac{\Phi}{2\pi}| + 
1, \frac{m \omega_{B}}{2 \hbar} \rho^{2} \right)     ,
\end{eqnarray}
\noindent
where $\omega_{B} = \frac{qB_{o}}{mc}$. Then, the energy is given by

\begin{eqnarray}
E&=& \hbar \omega_{B} \left\{ n + \frac{| \ell - \beta k +
\frac{\Phi}{2\pi} |}{2} - \frac{(\ell - \beta k+ 
\frac{\Phi}{2\pi} )}{2} + \frac{1}{2} \right\} \nonumber \\  
&&+ \frac{\hbar^{2} k^{2}}{2m}
\end{eqnarray}
\noindent
with $n = 0,1,2...$, and the eigenfunction  by

\begin{eqnarray}
\psi ( \rho , \phi , z) &=& C \; e^{ikz} e^{i \ell \phi} \exp \left\{ 
\frac{m \omega_{B} \rho^{2}}{4 \hbar} \right\} \rho^{| \ell - 
\beta k + \frac{\Phi}{2\pi} |} \times \nonumber \\
&& F \left( - n , | \ell - \beta k+ \frac{\Phi}{2\pi}| + 1, 
\frac{m \omega_{B}}{2 \hbar} \rho^{2} \right)  .
\end{eqnarray}

Note that, the energy levels have the  infinite degeneracy of classical 
Landau levels broken by coupling the torsion $\beta$ and the internal magnetic 
flux $\Phi$ of the defect  with the angular momentum $\ell$. In this case, the 
degeneracy of the levels is more strongly broken than in the case of a 
disclination. In the case of Landau levels, the coupling of angular 
momentum with parameter $\frac{1}{\alpha}$ is multiplicative and in the 
present case the coupling with parameter $\beta$ is additive. This fact 
is responsible by the strong break of degeneracy  of Landau levels in the 
presence of a screw dislocation. A special case will be considered if we 
observe that for some values of the magnetic field the influence of 
elastic properties of the medium can be canceled by an internal magnetic field
such that

\begin{eqnarray}
\label{cond}
\Phi=2\pi\beta k .
\end{eqnarray}

\renewcommand{\thesection}      {\Roman{section}}

\section{Landau levels in the Presence of a Dispiration}

In this section we analyze the Landau levels problem in the presence of 
dispiration. In general, this defect corresponds to singular curvature and 
torsion  along the defect line \cite{KV}. We consider an infinitely 
long linear dispiration oriented along the $z$-axis. The three-dimensional 
geometry of the medium is characterized by nontrivial torsion and 
curvature which are identified with the surface density of the Burgers and 
Frank vectors, respectively, in the classical theory of 
elasticity. In this way, the Burgers vectors can be viewed as a flux of 
torsion and the Frank vector as a flux of curvature. This defect is 
described by the following metric

\begin{equation}
ds^2=(dz+\beta d\phi)^2+\alpha^2\rho^2d\phi^2+d\rho^2 \label{24}
\end{equation}
\noindent
where  $\beta=\frac{b}{2\pi}$ and $\alpha=(1+\frac{\lambda}{2\pi})$. This 
metric is equivalent to the following construction: removal $(\alpha < 1)$ or 
insertion $(\alpha >1)$ of a wedge of material of dihedral angle 
$ \lambda =2 \pi( \alpha - 1)$ followed by a
translation of the  lips with respect to each other of $b$ along the 
$z$-direction, according to the Volterra process\cite{Kle}, 
$b$ being the Burgers vector. The torsion two-form is the same as that of a 
screw dislocation and the curvature tensor the same of a disclination. This 
defect has two conical singularities, one due to torsion and the other to 
curvature. The vector potential that produces the uniform magnetic field in 
this metric is given by

\begin{equation}
A(\rho)=\frac{B\rho}{2\alpha}\ .  \label{f25}
\end{equation}

For a quasi-particle of effective mass $m$ in the metric of the  
dispiration, the Hamiltonian  is 

\begin{eqnarray}
\hat{H}&=& -\frac{\hbar^{2}}{2}\left\{ \partial_{z}^{2}+
\frac{1}{\rho}\partial_{\rho}(\rho \partial_{\rho})+
\frac{1}{\alpha^{2}\rho^{2}}(\partial_{\phi}-\beta \partial
_{z})^{2}\right\}\nonumber \\ [0.8cm]
&& +\frac{iqB\hbar}{2mc\alpha}(\partial_{\phi}-\beta \partial_{z})+
\frac{q^2B^2\rho^2}{8mc^2\alpha^2}\ .
\label{f26}
\end{eqnarray}
\noindent
The Schr\"{o}dinger equation corresponding to this case can be solved 
using the ansatz

\begin{equation}
\psi(\phi,\rho,z)=Ce^{il\phi}e^{ikz}R(\rho)\ ,  \label{f18a}
\end{equation}

\noindent
where $C$ is a normalization constant. Substituting this 
wave function into the Schr\"{o}dinger equation we obtain the following 
radial equation

\begin{eqnarray}
& & \frac{1}{\rho} \partial_{ \rho }( \rho \partial_{\rho} R(\rho)) - 
\frac{1}{\rho^{2}\alpha^{2}} ( \ell - \beta k)^{2} R( \rho ) - 
\frac{q^{2}B^{2}\rho^{2}}{4 \hbar^{2}c^{2}\alpha^{2}}R(\rho) \nonumber \\
& & - \frac{qB}{2 \hbar c\alpha} ( \ell - \beta k) R( \rho ) - 
k^{2} R(\rho ) + \varepsilon R(\rho) = 0 \label{se2},
\end{eqnarray}
with $\varepsilon = \frac{2mE}{\hbar^{2}}$. Now, by using the change of 
variables $\sigma = \frac{\rho^{2}}{2}$, eq.(\ref{se2}) is transformed into 

\begin{eqnarray}
&&\sigma^{2} \frac{d^{2}R}{d \sigma^{2}} + \sigma \frac{dR}
{d \sigma} \nonumber \\
&+& \left\{ \frac{A}{2}\sigma - \frac{q^{2} B^{2} \sigma^{2}}
{4 \hbar^{2} c^{2}} - \frac{( \ell - \beta )^{2}}{4} \right\} R(\sigma)=0 \  ,
\end{eqnarray}
\noindent
where $ A = \varepsilon - k^{2} - \frac{qB}{2\hbar c\alpha}(\ell - 
\beta k) $. The solution of this equation is given in terms of the 
confluent hypergeometric function $F$ 

\begin{eqnarray}
R( \rho )&=& C \exp \left\{ - \frac{m \omega_{B}}{4 \hbar\alpha} 
\rho^{2} \right\} \rho^{\frac{| \ell - \beta k|}{\alpha}} \times \nonumber \\
&& F \left( - n \; , \; \frac{| \ell - \beta k |}{\alpha} + 1, 
\frac{m \omega_{B}}{2 \hbar\alpha^{2}} \rho^{2} \right)     ,
\end{eqnarray}
\noindent
where $\omega_{B} = \frac{qB_{o}}{mc}$. Then, the energy  is 

\begin{equation}
E=\frac{\hbar w_B}{\alpha}
\left\{n+\frac{|l-\beta k|}{2\alpha}-\frac{(l-\beta k)}{2\alpha}+
\frac{1}{2}\right\}+\frac{\hbar k^2}{2m}\ , \label{f27}
\end{equation}
\noindent
where  $n=0,1,2,\dots$, and  the eigenfunction is given by the following 
expression

\begin{eqnarray}
\psi(\rho,\phi,z)&=&  Ce^{ikz}e^{i\ell \phi}
\exp\left[\frac{m\omega_{B}\rho^2}{4\hbar\alpha}\right]
\rho^{\frac{|l-\beta k|}{\alpha}}\times \nonumber \\ [0.5cm]
&& F\left(-n,\frac{|l-\beta k|}{\alpha}+1,\ 
\frac{m\omega_{B}}{2\hbar\alpha^2}\rho^{2}\right)\ .
\label{f28}
\end{eqnarray}

From this expression we note that the presence of the defect breaks the 
degeneracy of the energy levels. Comparing this result with the one of the
previous section, we conclude that in this case the degeneracy is strongly 
broken due to the influence of both parameters $\alpha$ and $\beta$.
Note that if we take $\alpha=0$, we get the results for a 
dislocation\cite{Fur1} and for $\beta=0$, we get similar results for a 
disclination\cite{F}.

\renewcommand{\thesection}      {\Roman{section}}

\section{Landau levels in Kaluza-Klein theory}

In this section we consider the non-relativistic problem concerning 
Landau levels in the presence of a dispiration in the framework of
Kaluza Klein theory. In this theory the electromagnetic field is introduced 
in a geometrical way as an extra dimension and the standard line 
element is given by 

\begin{eqnarray}
\label{Kl}
ds^{2}= g_{\mu \nu}dx^{\mu}dx^{\nu} + g_{55}(dx-KA_{\mu}dx^{\mu})^{2},
\end{eqnarray}
where $g_{55}$ is Kaluza-Klein scalar potential which will be considered
as $g_{55}=1$, K is the Kaluza constant and $x$ is the fifth cooedinate. 
In this context, the five-dimensional metric that corresponds to a uniform 
magnetic field in the presence of a dispiration is 

\begin{eqnarray}
ds^{2} &=& d\rho^{2} + (dz^{2}+ \beta d\phi)^{2}+ \alpha^{2} \rho^{2}  
d \phi^{2}\nonumber \\
&+&
\left( dx - \frac{B_{0}\rho ^{2}}{2} d \phi \right)^{2} , \label{c1}
\end{eqnarray}
where $A_{\phi} = B_{o}\rho/(2 \alpha)$
is the vector potential  and the magnetic field is $B^{z} = B_{0}$. The 
metric contains the elastic deformation caused by the defect in the medium and
has the coupling between the fifth coordinate and the $\phi$ coordinate.

Writing Schr\"odinger equation given by eq.(\ref{se}) in the space 
with metric (\ref{c1}), we get

\begin{eqnarray}
&-& \frac{1}{2m} 
\partial^{2}_{z}  +  \frac{1}{\rho} \partial_{\rho} ( \rho \partial \rho) +
\partial^{2}_{x} \nonumber \\
& + & \frac{1}{\alpha^{2} \rho^{2}} 
\left( \partial_{\phi}- \beta\partial_{z} - \frac{B_{0} \rho^{2}}{2} 
\partial_{x} \right)^{2} \psi
= i \frac{\partial \psi}{\partial t} . \label{c3}
\end{eqnarray}

Now, let us consider the ansatz

\begin{eqnarray}
\label{anz}
\Psi(x,t,\rho,\phi,z)= exp[-iEt + ikz +iQx +il\phi]R(\rho).
\end{eqnarray}

Substituting eq.(\ref{anz}) into eq.(\ref{c3}), we obtain the 
following radial equation

\begin{eqnarray}
\label{rad}
&&  \left\{ \frac{1}{\rho}\frac{d}{d\rho} +
\frac{1}{\alpha^{2}\rho^{2}}[\ell - \beta k -\frac{B_{0}\rho^{2}}{2}]^{2} 
\right. \nonumber \\
&+& \left.  (2mE -k^{2} -Q^{2})\right\}R(\rho)=0.
\end{eqnarray}

In this case we adopt the same procedure that we have adopted in previous
sections. Let us introduce a new variable $\eta$ such that 
$\eta = \frac{\rho^{2}}{2}$. Then, equation (\ref{rad}) turns into

\begin{eqnarray}
\label{eta}
&&\eta^{2}\frac{d^{2}R(\eta)}{d\eta^{2}} + \eta\frac{dR(\eta)}{d\eta} - 
\frac{(\ell-\beta k)}{4\alpha^{2}} + \frac{B_{0}Q(\ell-\beta k)}
{2\alpha^{2}}R(\eta)\nonumber\\
&&-\frac{B_{0}^{2}Q^{2}\eta^{2}}{4\alpha^{2}}R(\eta)+ 
(A^{2}-k^{2}-Q^{2})R(\eta)=0 ,
\end{eqnarray}
\noindent
where $A^{2}=2mE$. The solution of this equation is the confluent 
hypergeometric function. Therefore, the wave function is given by

\begin{eqnarray}
&&\Psi (t, \rho , \phi , z , x)  = 
C_{n\ell}e^{-iEt+ikz+iQx+i\ell\phi}\times \nonumber \\
&& e^{-\frac{B_{0}Q\rho^{2}}{4\alpha}}\rho^{|\ell-\beta k|}F(-n,\frac{|\ell- 
\beta k|}
{\alpha}+1,\frac{\rho^{2}}{2}),
\end{eqnarray}
and the eigenvalues are 

\begin{eqnarray}
E = \frac{B_{0}Q}{m\alpha} \left( n + \frac{| \ell-\beta k |}{2 \alpha} - 
\frac{\ell-\beta k}{2
\alpha} + \frac{1}{2} \right) + \frac{K^{2}}{2m} + \frac{Q^{2}}{2m} .
\label{kalu}
\end{eqnarray}

This result is in agreement with our earlier work~\cite{F} where
the magnetic field was introduced by minimal coupling. Note that in the limit
 $\alpha \to 1$ we obtain the result of Landau levels in the presence of a
screw dislocation. For $\beta=0 $, we obtain the results corresponding to
a dislocation. If we define $\omega=\frac{B_{0}Q}{m}$, the eigenvalues 
(\ref{kalu}) are almost the same of the dispiration, which were obtained
using the minimal coupling. The unique difference is 
the extra term $\frac{Q^{2}}{2m}$ in eq.(\ref{kalu}), which is the 
kinetic energy associated with the fifth dimension, with quantum number 
given by the charge of the particle.

\renewcommand{\thesection}      {\Roman{section}}

\section{Concluding Remarks}

In this paper we study Landau levels in the presence of a class of 
topological defects. We demonstrate that the presence of topological 
defects breaks the infinite degeneracy of Landau levels, due to the
unusual boundary conditions imposed by the defects. The advantage of this
geometrical method to treat these problems are the easy and  exact 
calculations employed, in contrast with the  theory of elasticity, in which 
the exact solutions of the simple problem in general are not possible. 
Generally, the determination of Landau levels in the framework of the 
theory of elasticity are taken  using perturbation methods to solve 
Schr\"odinger equation\cite{KF,KT}.The degeneracy of the Landau levels in 
the presence of a continuous distribution of disclinations is smoothly 
broken. The coupling of the curvature of this conical defect to the angular 
momentum gives the possibility of breaking the degeneracy for some specific 
values of $\alpha$. In the case of a screw dislocation, the coupling 
of Burgers vector $\beta=\frac{b}{2\pi}$ to the angular momentum $\ell$ is 
responsible for the breaking of degeneracy of Landau levels. In the case 
of a  dislocation, we see that the torsion affects more strongly the 
Landau levels. For a dispiration, a defect that carries both curvature 
and torsion, we can conclude that the degeneracy of the Landau levels
is more strongly broken than in  all previous cases. The combination of the 
couplings of torsion $\beta$ and curvature $\alpha$ is responsible for this 
effects on the energy levels.  The use of Kaluza-Klein theory is a powerful 
geometric method to study Landau levels. In this theory, we introduced the 
elastic deformation of a media with the homogeneous magnetic field included 
in the metric that describes the dispirated media. In this approach, 
we obtain the results concerning Landau levels in an elegant way.

\noindent

{\bf Acknowledgement}

This work was partially supported by CNPq and FINEP.

\end{document}